\documentclass[draft]{amsart}
\usepackage{amssymb}

\newtheorem{theorem}{Theorem}[section]
\newtheorem{lemma}[theorem]{Lemma}

\theoremstyle{definition}
\newtheorem{definition}{Definition}

\begin{document}
 \renewcommand{\theequation}{\thesection.\arabic{equation}}

\title[FOUNDATIONS OF STATISTICAL MECHANICS AND THEORY OF PHASE
TRANSITIONS]{Foundations of statistical mechanics and theory of
phase transitions} \author{E.D. Belokolos} \address{Department of
Theoretical Physics \\ Institute of Magnetism, National Academy of
Sciences of Ukraine \\ 36--b Vernadsky str., Kiev-680, 252142,
Ukraine \\ tel: +38 044 444 3274; Fax: +38 044 444 1020, e-mail:
belokolos@guukr.freenet.kiev.ua} \date{\today} \begin{abstract} A
new formulation of statistical mechanics is put forward
ac\-cor\-ding to which a random variable characterizing a
macroscopic body is postulated to be infinitely divisible. It
leads to a parametric representation of partition function of an
arbitrary macroscopic body, a possibility to describe a
mac\-ro\-sco\-pic body under excitation by a gas of some
elementary quasiparticles etc. A phase transition is defined as
such a state of a macroscopic body that its random variable is
stable in sense of L\'evy.  From this definition it follows by
deduction all general properties of phase transitions:  existence
of the renormalization semigroup, the singularity classification
for thermodynamic functions, the phase transition universality and
universality classes. On this basis we has also built a
$2$-parameter scaling theory of phase transitions, a thermodynamic
function for the Ising model etc.  \end{abstract} \maketitle
\tableofcontents \section*{Introduction}

A creation of phase transition theory is one of the most
important problem of statistical mechanics. While we have reached
big success last years we have no still solution of the
problem\cite {ma78}, \cite{pp82}.

In this paper we propose a new approach to this problem which is
based on analysis of a random variable characterizing
a macroscopic body. According to traditional exposition of
statistical mechanics this random variable is to be a
me\-cha\-ni\-cal additive integral of motion. Generalizing this
statement we postulate that random variables in statistical
mechanics must be infinitely divisible.  Using a canonical
representation by L\'evy--Khintchine for a characteristic
function of infinite divisible random variable  we have got a
parametric representation of partition function of any macroscopic
body and therefore a possibility to analyze its
ther\-mo\-dy\-na\-mic properties.  Since a generalized Poisson
random variable, which from physical point of view is nothing else
than the Boltzmann ideal gas random variable, can approximate
infinite divisible random variable as good as we wish we can
explain why it is possible to imagine any excited macroscopic body
as a gas of some quasiparticles or elementary excitations. A
general representation of a set of random variables characterizing
macroscopic bodies in statistical mechanics follows from the fact
that a set of infinite divisible random variables is a convex set
having the normal and Poisson random variables as extreme points.

Analysis of modern physical representations on a nature of phase
transitions leads to an idea that the phase transition is such a
state of a macroscopic body that its random variable is stable in
sense of L\'evy. Taking this as a definition we has become able to
deduce all well known properties of phase transitions. The
stability property itself uses in its definition as a matter of
fact a notion of renormalization semigroup which is well familiar
in statistical mechanics. A number of finite moments for a stable
random variable or, which is the same, a number of finite
derivatives of a partition function for a macroscopic body, is no
more than the so called characteristic exponent $ \alpha, 0<
\alpha<2$, which plays role of a critical index for the phase
transition. It means that in the nature there exist phase
transitions only of the first and second order. Using a notion of
attraction domain for the stable random variable we has been able
to explain a meaning of a phase transition universality and define
universality classes.  Distribution functions of random variables
describing macroscopic bodies in a region of phase transition are
appeared to be also universal and are expressed in terms of the Fox
functions.

We have applied the above theory to a case when macroscopic body
is cha\-rac\-te\-ri\-zed by a $2$-dimensional random variable
(e.g.  an energy, an order parameter).  We have succeed to
reproduce all results of a $2$-parametric scaling theory and
express all critical indexes in terms of appropriate
characteristic exponents.  In a case when one of characteristic
exponents is equal to 1 the theory leads to well known results of
the exactly solvable Ising model with field.

The paper is organized as follows. In chapter 1 we postulate that
the random variable characterizing  any macroscopic body is
infinite divisible and consider dif\-fe\-rent consequences of this
postulate.  In chapter 2 we give the definition of phase
transition and then reproduce all characteristic properties of
phase transitions at first for an elementary case of
the $1$-dimensional random variable and further for a more general
case of the many-dimensional random variable. In chapter
3 we consider an example of application of general theory to a
particular case of $2$-dimensional random variable and as a result
deduce a $2$-parametric scaling theory of phase transitions.
At last we conclude the paper with some remarks about general
meaning of our results.

\section{Infinitely divisible random variable describes
macroscopic body in statistical mechanics}

A statistical mechanics studies the properties of macroscopic
bodies which con\-sist, according to definition, of an immense
number (e.g.  the Avogadro number) of some identical particles
(e.g.  atoms, molecules).  As result of that the main laws of
statistical mechanics are laws of a collective, a crowd and every
macroscopic body is described by a random variable. The random
variable $X$ is certainly to be additive and an appropriate
probability measure $\mu(x)$ is surely to be multiplicative
relative to a convolution $\ast$, i.e. if we divide a macroscopic
body into $n$ independent subbodies then \begin{eqnarray}
X=X_{1}+\cdots+X_{n},\\ \mu (x)=\mu (x_{1})\ast\cdots\ast
\mu(x_{n}). \end{eqnarray} The described above scheme of partition
and successive construction of a mac\-ro\-sco\-pic body out of its
subparts (we call it the algorithm of Democrites of Abdera) is
used constantly in statistical mechanics.  Of course we suppose
that such a partition is possible for every positive integer n. It
means that a following postulate is valid:

{\it a random variable corresponding to a macroscopic body is
infinitely divisible.}

This postulate generalizes and takes place of a similar statement
used in the traditional exposition of statistical mechanics due to
Gibbs according to which the random variable of macroscopic body
must be a mechanical additive integral of motion. We consider this
replacement reasonable since in our opinion we have to formulate
the general principles of statistical mechanics without any
reference to other parts of physics (e.g. the classical or quantum
mechanics). In essence this postulate means that if the atom or
molecule dimensions are negligible we may apply the Democrites
algorithm of decomposition of  a macroscopic body into subbodies
endlessly.

We can formulate this postulate as

\begin{definition} {\it A macroscopic body in statistical
mechanics is the physical structure which is
described by an infinitely divisible random variable.}
\end{definition}

Further we are going to study random variables taking values in
${\mathbb R}^k$. We shall consider every random variable to be
full i.e. we shall suppose that a support of an appropriate
probability measure do not belongs to a proper hyperplane of
${\mathbb R}^k$. By changing an origin of coordinates and a unit
of measurement we undergo a random variable $X$ to an affine
transformation $X\mapsto AX+b$, where $A$ is a nondegenerate linear
transformation and $b$ is a vector. Random variables are said to be
of the same type if they are connected by an affine
transformation. Since this relation is reflexive, symmetric and
transitive a set of random variables is divided into mutually
disjoint types.

A probability measure of a macroscopic body is known to have
such a form (Gibbs, J.W., 1902) \cite{gib02}
\begin{eqnarray} \label{gib}
\mu (dx)=Z^{-1} (u)\exp(-<u,x>)D(x)dx,
\end{eqnarray}
where $<u,x>$ is a linear functional on a space of random
variables, $D(x)$ is a density of states of the macroscopic body
for a given value x of a random variable and \begin{eqnarray}
Z(u)=\int\exp(-<u,x>)D(x)dx
\end{eqnarray}
is a partition function.

A logarithm of a partition function is well known to be
a generating function for cumulants of the random variable and
besides of that defines thermodynamic functions of the macroscopic
body. If for example the additive random variable is an energy then
\begin{eqnarray}
F(u)=-u^{-1}\ln Z(u)
\end{eqnarray}
is the Helmholtz free energy and $u^{-1}$ is a temperature.

Since it is better to describe properties of infinitely divisible
random variables in the language of characteristic functions we
state first of all a connection between the characteristic function
\begin{equation} \label{eps}
\hat\mu (y)=M\exp(i\langle y,x\rangle )=\int\exp(i\langle y,x\rangle )\mu (dx)
\end{equation}
and the partition function $Z(u)$.

\begin{lemma}
\begin{equation}
\ln \hat\mu (y,u)=\mathrm{ln} Z(u-iy)-\mathrm{ln} Z(u).
\end{equation}
\end{lemma}

\begin{proof} Proof consists of substitution of the expression for
macroscopic  body pro\-ba\-bi\-li\-ty measure (\ref{gib}) in
the definition of characteristic function (\ref{eps}).\end{proof}

It follows from the lemma that the characteristic function
$\hat\mu (y,u)$ after analytical continuation in the variable
$y$ to the complex space ${\mathbb C}^k$ has such an expression
\begin{equation} \label{dra} \ln\hat\mu (iv,a)=\ln Z(u+v)-\ln
Z(u).  \end{equation}

If we do not want to use results of the traditional approach to
statistical me\-cha\-nics we should consider the lemma and
therefore the formula (\ref {dra}) as a postulate.

According to L\'evy P. and Khintchine A.J. \cite {gak68} the
logarithm of characteristic function of an infinitely divisible
random variable has such a canonical form \begin{eqnarray}
\label{ura} \ln \hat\mu (y)=i\langle y,a\rangle -(1/2)\langle
y,Ry\rangle +\nonumber\\ +\int_{{\mathbb R}^k \setminus \lbrace 0
\rbrace} \left( \exp{(i\langle y,x\rangle )}-1-\frac{i\langle
y,x\rangle }{1+\Vert x \Vert ^{2}}\right) M(dx), \end{eqnarray}
where $a \in {\mathbb R}^k$, $R$ is symmetric non-negative bounded
operator in ${\mathbb R}^k$ and a so called L\'evy measure $M$
satisfies a condition \begin{displaymath} \int_{{\mathbb R}^k}
(1\bigwedge\Vert x \Vert ^{2}) M(dx)<+ \infty.  \end{displaymath}

Since the canonical form is unique we may wright down an
infinitely divisible probability measure in a following way
\begin{equation} \mu = [a,R,M].
\end{equation}

The canonical form of the logarithm of characteristic function for
the infinitely divisible random variable is valid under analytical
continuation in the variable $y$ to any containing the point 0
convex tubular domain $G \subset {\mathbb C}^k$ and an
appropriate integral in this formula converges absolutely and
uniformly in any compact $K \subset G$. Let us remind to this end
that the domain $G \subset {\mathbb C}^k$ is named as the
tubular one with base $B \subset {\mathbb R}^k$ if it is of the
form \begin{displaymath} G=\lbrace z \in {\mathbb C}^k \vert \Im
z \in B \rbrace.  \end{displaymath}

Now we can prove a following

\begin{theorem} \label{rum} A logarithm of the partition function
of any macroscopic body has such a canonical form
\begin{eqnarray} \ln Z(u+v)=\ln Z(u)-\langle v,a\rangle
+(1/2)\langle v,Rv\rangle +\nonumber\\ +\int_{{\mathbb
R}^k\setminus \lbrace 0 \rbrace} \left( \exp{(-\langle
v,x\rangle )}-1+\frac{\langle v,x\rangle }{1+\Vert x \Vert
^{2}}\right) M(dx), \end{eqnarray} where $a\in {\mathbb R}^k$,
$R$ is symmetric non-negative bounded operator in ${\mathbb
R}^k$ and $M$ is a L\'evy measure.\end{theorem}

\begin{proof} It follows from formulae (\ref
{dra}) and (\ref {ura}).\end{proof}

This parametric expression for the logarithm of partition function
of macroscopic body (and therefore for its thermodynamic
functions) must be of extreme im\-por\-tance in statistical
mechanics.  It is obvious that different classes of parameters
must describe different classes of macroscopic bodies and even
different classes of mic\-ro\-sco\-pic theories that lay in their
foundations. Let us remind in this connection that the quantum
mechanics has appeared for the first time as a new kind of
distributions for a special case of a black radiation. In next
section of the paper we use the canonical representation of
partition function in order to build up the phase transition
theory.

An infinite divisibility of random variable characterizing a
macroscopic body is tightly connected to a scheme of decomposition
of the macroscopic body into independent subbodies. Let us
consider to this end a sequence of decompositions of the
macroscopic body into subbodies when at a step $n$ we partition the
body into $k(n)$ subbodies and denote a random variable of the
$j$--th subbody as $X_{n,j}, 1 \leq j \leq k(n)$. Thus we
get a sequence of series of independent random variables
$X_{n,j}, 1 \leq j \leq k(n)$. Further we shall suppose these
random variables $X_{n,j},  1 \leq j\leq k(n)$  to be infinitely
small, i.e.  for any $ \epsilon > 0 $ such a condition will be
valid:  \begin{equation} \label{bur} \lim_{n \to \infty} \qquad
\sup_{1 \leq j \leq k(n)} P \lbrace \vert X_{n,j}\vert > \epsilon
\rbrace =0 \end{equation}

Then a limit distribution of a sum
\begin{displaymath}
S_{n}= \sum_{j=1}^{k(n)} X_{n,j}
\end{displaymath}
exists if and only if the random variable is infinitely divisible.

Let us discuss now a physical interpretation for some classes
of infinitely divisible random variables.

A random variable with normal distribution is well known to be
infinitely di\-vi\-sib\-le. It appears as a limit in the
sequences of series of independent random variables when and only
when for any $\epsilon > 0$ is valid a stronger  condition
\begin{equation}  \label{mur}
\lim_{n \rightarrow \infty} \sum_{j=1}^{k(n)} P
\lbrace \vert X_{n,j}\vert > \epsilon \rbrace =0
\end{equation}

We designate conditions (\ref{bur}) and (\ref{mur}) as the
conditions of weak and strong fluctuations appropriately.
Using this terminology we may state that the random variables with
normal distribution should characterize weak fluctuations of
the thermodynamic quantities in a universal way. This is indeed
the case in statistical mechanics.

A random variable with generalized Poisson distribution is also
known to be infinitely divisible. A probability measure for this
case has the following form
\begin{equation}
\mu = \exp {\left(- \nu({\mathbb R}^k)\right)}
\sum_{n=0}^{\infty} \nu^{\ast n}/ n!,  \end{equation} where $\nu$
means some single-particle measure. In physics we denote the
generalized Poisson distribution  as the Boltzmann  distribution
which describes an ideal gas of non-interacting particles of
classical mechanics. We should remark here that the factor $n!$ in
this formula needs special explanation in a traditional
derivation of the Boltzmann distribution from the Gibbs
one by a necessity to take in account the identity of particles or
to ensure the additivity of thermodynamic functions.
In our approach all these properties follows from the infinite
divisibility of the Poisson distribution.

Let us consider now some consequences of the infinite divisibility
of random variables which characterize macroscopic bodies in
statistical mechanics.

First of all we should stress a fundamental importance of the
Poisson distributions in a theory of infinitely divisible
distributions: any infinitely divisible distribution is a limit of
the Poisson distributions. This remarkable property of the Poisson
distributions means that a following theorem is valid.

\begin{theorem} Any excitation of macroscopic body is possible
to present as an ideal gas of some elementary
excitations or quasiparticles with properties depending
on a type of the macroscopic body.\end{theorem}

Thus we can give for the first time a natural explanation of the
fact well known in physics on an existence in macroscopic bodies
different kinds of quasiparticles such as phonons, plasmons,
excitons, polarons, magnons etc. This fact as we have seen is
just a simple consequence of the infinite divisibility of a random
variable characterizing a macroscopic body.

Of course if we need to use the quantum mechanics in order to
describe physical properties of a macroscopic body we must change
the Poisson distribution by the Bose and Fermi distributions which
are also infinitely divisible as it easy to show.

According to the canonical representation (\ref {ura}) a set of
infinitely divisible pro\-ba\-bi\-li\-ty measures is a convex hull
of the normal and Poisson measures (Johanssen, S. 1966) \cite
{joh66}.  This mathematical fact is based on a fundamental
theorem by Krein M.G., Milman D.P. \cite{km40} and has also
remarkable physical interpretation:  pro\-ba\-bi\-li\-ty measures
of macroscopic bodies form a convex hull of the probability
measures of fluctuations and ideal gases.

Thus in this section we has given a new formulation of the
statistical mechanics which is based on the infinite divisibility
of random variables that characterize macroscopic bodies. In a
frame of our approach we do not need to use any notions of
microscopic theories such as e.g. the hamiltonian, moreover our
approach form natural external bounds for any possible microscopic
theories. We have been able to wright down the partition
function for any macroscopic body and explain a possibility to
describe any excitation of it by an ideal gas of some
quasiparticles. It seems that new formulation is simpler and
deeper than usual one and more effective especially in studies of
general questions of statistical mechanics. We demonstrate it in
the next section.

\section{Stable random variable describes macroscopic body in a
state of phase transition.}

We have shown above that a random variable of any macroscopic
body is infinite divisible. Changing the parameter $u \in {\mathbb
R}^k$ we can change essentially properties of the random
variable although it continues to be infinitely divisible all the
time.  In particular this random variable can become stable in
sense of L\'evy, P. (L\'evy, P. 1924) \cite
{lev54}.  Analyzing the modern physical representations on phase
transitions we have come to a conclusion that the following
definition is valid.

\begin{definition} {\it The phase transition is such a state
of macroscopic body  when the appropriate random variable is
stable.} \end{definition}

From this definition we deduce further all well known basic
properties of phase transitions. But at first in  order  to make
our exposition as simple as possible and concentrate on main ideas
of the work we consider the case $k=1$, i.e. we shall suppose that
all random variables under consideration take their values in
${\mathbb R}$.

{\em á. Renormalization  semigroup.}

A stability of the random variable characterizing a
macroscopic body under phase transition means the following: if we
split the macroscopic body into $n$ subbodies with independent and
equally distributed random variables $X_{k}, 1 \leq k \leq n$
then we can find always such real numbers $a_{n} > 0$ and $b_{n}$
that
\begin{equation} \label{fig}
\sum_{k=1}^{n} X_{k}=a_{n} X+ {b_{n}}.
\end{equation}
Here the symbol $=$ means an equality of random variables
in their distributions. It follows easily from this equality that
\begin{displaymath}
a_{n}=n^{1/ \alpha}, \qquad 0 < \alpha \leq 2.
\end{displaymath}

A quantity $\alpha$ is usually called the characteristic
exponent of a stable random variable. If $\alpha =0$ then the
appropriate random variable $X$ is degenerate i.e. it takes with a
probability 1 only  one value. If $\alpha =2$ then the random
variable has normal distribution. A stable random variable $X$
is called  strictly stable if $b_{n}=0$. If a stable random
variable $X$ has the characteristic exponent $\alpha \not=1$ then
there exists such a constant $c$ that the random variable $X-c$
is strictly stable.

The condition $(\ref{fig})$ which characterizes the stability
property of a random va\-ri\-ab\-le is defined only for natural
numbers $n$ but we can generalize it easily to all real positive
numbers $t$. Being expressed in terms of a probability measure
$\mu$ the stability condition $(\ref{fig})$ takes such a form:
\begin{equation} \label{tyu} \mu ^{t}=t^{1/ \alpha}\mu \ast \delta
(b_{t}), \qquad t > 0.  \end{equation} Here the measure $\mu^{t}$
means the $t$-th convolution power of the measure $\mu$, the
measure $t^{1/\alpha} \mu $ means the probability
measure for the random variable $t^{1/\alpha} X $, i.e.
$t^{1/\alpha} \mu (E)= \mu (t^{-1/\alpha} E) $ for any Borel set
$E$, and $\delta (b)$ means the Dirac measure in a point $b$.

According to the definition $(\ref{tyu})$ a stable probability
measure is invariant under a transformation semigroup with
elements consisting of a convolution power of the measure and
a subsequent its affine transformation. We designate this semigroup
as the renormalization one since it is a proper
mathematical generalization of that used now widely in a
modern theory of phase transitions \cite{ma78}, \cite{pp82}. Thus
the stability of a probability measure for a case of phase
transition means just its invariance with respect to the
renormalization semigroup. In other words the probability
measure under phase transition is a fixed point of the
renormalization semigroup transformations. We want to emphasize
here that we have introduced the notion of renormalization
semigroup without any reference to the order of phase transition
and therefore it is valid for any type of transition. And it is
very essential difference of our renormalization semigroup and
that used now in the theory of phase transition.

{\em B. Singularities of thermodynamic functions. Classification
of phase transitions.}

A principal characteristic of a stable probability measure is
the characteristic exponent $\alpha $. If $\alpha =2$ then
the stable measure is normal and therefore it has moments of all
order. If $0 < \alpha <2 $  then the stable measure has finite
moments only of the order $\delta $ where $0< \delta
<\alpha $. The existence of moments of the order $\delta $ is
known to lead to the existence of derivatives of the order
$\delta $ for the logarithm of characteristic function what is
equivalent according to the lemma proven above to the existence of
derivatives of the order $\delta $ for the logarithm of partition
function or the thermodynamic function of the macroscopic body.
Thus these arguments proves that the thermodynamic functions of
macroscopic body under phase transition have derivatives only of
the order $\delta $ where $0< \delta <\alpha $ and the
characteristic exponent $\alpha $ plays role of the critical index
for the phase transition. Moreover the thermodynamic functions may
have only the power singularities if the characteristic exponent
$0 < \alpha < 2, \ \alpha \not= 1$ and may have only the
logarithmic singularity if $\alpha = 1.$ We can formulate these
arguments more precise.

\begin {theorem} \label{gum} The logarithm of partition function
under phase transition has singularities of such a type:

for $0< \alpha <2, \qquad \alpha \not=1 $

\begin{eqnarray}
\ln Z(u+v) - \ln Z(u) &=&  c_{1} \alpha \Gamma (-\alpha)
v^{\alpha}, \qquad v>0 \\ &=& c_{2} \alpha \Gamma (-\alpha)
\vert v \vert ^{\alpha}; \qquad v<0 \end{eqnarray}

for $\alpha =1$

\begin{eqnarray}
\ln Z(u+v) - \ln Z(u) &=&  c_{1} v \ln v, \qquad v>0 \\
&=& c_{2} \vert v \vert \ln \vert v \vert, \qquad v<0.
\end{eqnarray}
\end{theorem}

\begin{proof} After substitution of the canonical expression for
infinitely divisible random variable (\ref {ura}) in the stability
condition (\ref {tyu}) it is easy to prove that the infinitely
divisible random variable is stable if and only if its
L\'evy measure $M$ in the canonical expression has such a
form:

\begin{eqnarray} \label{lev}
M(x)&=& -c_{1}x^{- \alpha}, \qquad x>0 \\
&=& c_{2}(-x)^{- \alpha}, \qquad x<0,
\end{eqnarray}
where $c_{1}>0, c_{2}>0, 0< \alpha < 2 $.

It means that the logarithm of characteristic function of the
stable random variable is a sum of the Fourier transformation

\begin{equation}
F[g](y) = \int g(x) \exp (i\langle  x,y\rangle) dx \nonumber
\end{equation}
of the distributions $x_{+}^{- \alpha - 1}, \quad x_{-}^{- \alpha
- 1}$:

\begin{displaymath}
\ln \hat \mu (y) = iya +  c_{1} \alpha F[x_{+}^{- \alpha - 1}\; ]
(y) +  c_{2} \alpha F[x_{-}^{- \alpha - 1}\; ] (y) .
\end{displaymath}

If $0< \alpha < 2, \alpha \not= 1$ then

\begin{displaymath}
\ln \hat \mu (y) = iya + \alpha \Gamma (- \alpha) \left( c_{1}
\exp (-i \pi \alpha / 2)(y + i0)^{\alpha}+ c_{2} \exp (i \pi
\alpha / 2) (y - i0)^{\alpha} \right).  \end{displaymath}

After analytical continuation of this expression from a real
line $y \in {\mathbb R}$ to a complex plane $z=y+iv \in {\mathbb
C}$ we get

\begin{eqnarray*}
\ln \hat \mu (z) &=& iza + c_{1} \alpha \Gamma
(-\alpha)(-iz)^{\alpha}, \qquad \Im z >0, \\   &=& iza + c_{2}
\alpha \Gamma (-\alpha) (iz)^{\alpha}, \qquad \Im z >0 .
\end{eqnarray*}

If $\alpha = 1$ then

\begin{displaymath} \ln \hat \mu (y) = iya -  c_{1}iy \ln
\left( -i \frac{y+i0}{\exp(1- \gamma)}\right)  +  c_{2} iy
\ln \left( i \frac{y-i0}{\exp(1- \gamma)} \right) ,
\end{displaymath} where $\gamma = - \Gamma'(1)$ is the Euler
constant. After analytical continuation of this expression from a
real line to a complex plane we get

\begin{eqnarray*} \ln \hat \mu
(z) &=& iza -  c_{1}iz \ln \left(- \frac{iz}{\exp(1 -
\gamma)}\right), \qquad \Im z>0, \\ &=& iza +  c_{2} iz \ln
\left(\frac{iz}{\exp (1- \gamma)}\right), \qquad \Im z < 0.
\end{eqnarray*}

Let us take $\Im z = \pm  \vert v \vert$ in accordance with
$\Im z >0, \Im z < 0 $. At last using the lemma established
earlier we prove the statement of the theorem. \end{proof}

According to our theorem the logarithm of partition function is
continuous function of  a parameter $v$. If $0<\alpha <1$ then
the first derivative of the logarithm of partition function has a
power singularity $v^{\alpha - 1}$. If $1<\alpha <2$ then the
first derivative of the logarithm of partition function is
continuous but the second one has a power singularity $v^{\alpha - 2}$.
In statistical mechanics we call a phase transition of the $n$-th
order if the first $n - 1$ derivatives of a
thermodynamic function are finite but the $n$-th derivative is
singular (Ehrenfest, P. 1933)\cite {ehr33}. In accordance with this
classification of phase transitions a range of characteristic
exponents $0 <\alpha < 1$ corresponds to phase transitions of the
first order and a range $1<\alpha < 2$ corresponds to phase
transitions of the second order. If $\alpha = 1$ then the
first derivative of the logarithm of partition function has
a singularity which is a sum of a jump and a logarithmic
singularity and the second derivative has a power singularity
$v^{-1}$. We may designate the last type of phase transitions
as the $\lambda$-point. According to our theory there are no any
phase transitions of the order more than 2.

Phase transitions of the first and second order differ
substantially not only quantitatively by the value of their
characteristic exponents $\alpha$ but also qualitatively.
According to the probability theory a contribution to the sum
of random variables $X_{1} + \ldots + X_{n}$ appearing in the
stability condition (\ref{fig}) is due to only one summand
provided that $0 <\alpha <1$ and is due to all summands
equally provided that $1 <\alpha <2$ (see for example
\cite{fel66}).  This mathematical result corresponds to the
point of view generally accepted in physics according to
which a phase transition of the first order is the result of local
germs appearing by fluctuations and a phase transition of the
second order takes place at once in a whole volume of macroscopic
body.

In general situation we can not predict what order of
phase transition should be. We have only one
exception: if a random variable characterizing the macroscopic
body is bounded then  the phase transition is necessarily of the
first order.  In fact a stable random variable $X$ with parameters
\begin{equation} \alpha \; \mbox{and} \; \beta =
\frac{c_{1}-c_{2}}{c_{1}+c_{2}} \nonumber
 \end{equation} is
bounded from the left (from the right) if and only if $\beta = 1
(\beta = -1)$ and $0<\alpha < 1$ (Esseen, C.--G. 1965)\cite{ess65}.
In this connection we should remark that an infinite divisible
random variable may not be bounded from two sides because
otherwise it becomes degenerated.

{\em ó. Universality of phase transitions. Universality classes.}

The thermodynamic function of nacroscopic body
depends in a general state  according to Theorem \ref{rum} on
real numbers $a, R$ and a L\'evy spectral measure
$M$  and in a state of phase transition according to Theorem
\ref{gum} only on four real numbers $\alpha , c_{1}, c_{2}, a$.
Therefore when the macroscopic body pass from an
arbitrary general state to a state of phase transition
we get essential contraction of a set of parameters necessary
to describe  this body. As a result of that an immense set of
macroscopic bodies being characterized by different parameters in
a general state may be described by identical values of a small
number of parameters in a state of phase transition. Physicists
designate this property of phase transitions as universality.

Now we consider the universality of phase transition in details.
Let $X_{1}, X_{2}, \ldots$ be a sequence of independent
random variables with a same distribution
function $F(x)$.  If there exist constants $A_{n}, b_{n},$ of a
such a type that distribution functions of the normalized sums
$S_{n}=A_{n}^{-1} (\sum_{k=1}^{n}X_{k} - b_{n})$ converge
to a distribution function $G(x)$  when $n \to \infty$
we say that the distribution function $G(x)$ attracts the
distribution function $F(x)$. Further we shall call a set of
all distribution functions $F(x)$ attracted to the distribution
function $G(x)$ as a domain of attraction of the distribution
function $G(x)$.  According to a probability theory the
stable distribution functions and only they have domains of
attractions. When we interpret a stable random variable as a
random variable of macroscopic body in a state of phase transition
then we should interpret an appropriate domain of attraction as a
class of universality of this phase transition.

According to a probability theory the normal distribution function
attracts the largest set of distribution functions while domains
of attraction of other stable distributions consist of those
distribution functions which are in some sense similar to these
stable distributions e.g. they are expressed by the same formulae
as these stable distributions where however constants have to be
changed by slowly changing functions (in sense by
Karamata, J. 1930)\cite{kar30}.  More precisely the distribution
function $F(x)$ belongs to the attraction domain of a stable
random variable with a cha\-rac\-te\-ris\-tic exponent $\alpha,
0<\alpha <2$ if and only if for any constant $k>0$
\begin{eqnarray*}
\frac{1 - F(x) + F(-x)}{1 - F(kx) + F(-kx)} \to k^{\alpha}
\qquad \mbox{for} \qquad x \to \infty;\\
\frac{F(-x)}{1 - F(x)} \to \frac{c_{1}}{c_{2}} \qquad \mbox{for}
\qquad x \to \infty.
\end{eqnarray*}

Under these conditions we may choose in the following way the
constants $A_{n},b_{n}$ characterizing the normalization and
centering of a random variable:  \begin{eqnarray*} &(\star)& \quad
\mbox{for} \quad 0<\alpha <1 \quad \lim_{n \to \infty} nh(A_{n}) =
1, \quad b_{n} = 0; \\ &(\star \star)& \quad \mbox{for} \quad
1<\alpha <2 \quad \lim_{n \to \infty} nh(A_{n}) = 1, \quad b_{n} =
na, \quad \mbox{for} \quad a = M X_{k};\\& (\star \star \star)&
\quad \mbox{for} \quad \alpha = 1 \quad \lim_{n \to \infty}
nh(A_{n}) = 1, \quad b_{n} = n A_{n} \int \frac{x}{x^{2} + A_{n}}
dF(x).  \end{eqnarray*}

Here $h(x) = 1 - F(x) + F(-x).$ Comparing expressions for the
quantity $b_{n}$ in cases $0 <\alpha < 1$ and $1 <\alpha < 2$
we can give another formulation for a qualitative difference
between phase transitions of the 1-st and 2-nd order: a mean value
of the random variable is not changed under phase transition of
the 1-st order and is changed under phase transition of
the 2-nd order.

For constructing of the phase diagrams it may appear useful a
notion of a domain of partial attraction. Let $X_{1}, X_{2}, \ldots$
be a sequence of independent random variables with the same
distribution function $F(x)$.  If there exist such a subsequence
of natural numbers ${j}$ and constants $A_{j}, b_{j},$ that
distribution functions of normalized sums
$S_{j}=A_{j}^{-1} (\sum_{k=1}^{j}X_{k} - b_{j})$ converge
to a distribution function $G(x)$ provided that $j\to \infty$
then we say that the distribution function $F(x)$
belongs to a domain of partial attraction of the
distribution function $G(x)$. Thus a difference between
domains of complete and partial attraction rests on a
difference between a sequence of all natural numbers and its
proper subsequence. It appears that every infinitely divisible
random variable (more precisely, the appropriate type) has a
domain of partial attraction (Khintchine, A.J. 1937)\cite{khi37}.
Moreover every random variable belongs to a domain of partial
attraction of only one type or a countable set of types or does
not belong to any domain of partial attraction (D${\ddot
o}$blin, W. 1940, \cite{dob40}; Gnedenko, B.V.1940,
\cite{gne40}).  If under changing of external parameters a
macroscopic body can undergo to different phase transitions then
the appropriate random va\-ri\-ab\-le must move from a domain of
attraction of one stable random variable to another one.  This
movement may take place only through domains of partial
attraction.

{\em D. Density of a stable statistical distribution
under phase transition. Rational and irrational critical indexes.}

According to probability theory the  stable probability measure
has density which is analytical function with well known
power series expansion. Further for simplicity of exposition we
shall discuss only strictly stable measure $\mu$ which canonical
re\-pre\-sen\-ta\-ti\-ons is defined by the L\'evy
spectral measure only.  Since the L\'evy measure of
stable probability measure is power function by the formula
(\ref{lev}) then except of the Fourier transformation of this
measure it is reasonable to consider also the Mellin
transformation of it \begin{equation} M[g](s) = M(s) =
\int_{0}^{\infty} g(x) x^{s - 1}dx. \nonumber \end{equation} It is
easy to show that the Mellin transformation of stable probability
measure has the following expression:  \begin{equation}
\label{lin} M(s|\alpha, \rho) = \frac{\Gamma(s - 1) \Gamma \left(1
+ 1 /\alpha - s/\alpha\right)}{\Gamma(\rho s - \rho)\Gamma(1 +
\rho - \rho s)} \end{equation}

Here we characterize a stable random measure by parameters
$\alpha, \rho$ instead of the parameters $\alpha, \beta$ used
earlier.  A connection between the parameters $\beta$ and $\rho$
has such a form:  \begin{equation} \beta = \cot (\pi \alpha / 2)
\tan (\pi \alpha (\rho - (1/2))) \nonumber \end{equation}

Now let us carry out an inverse Mellin transformation
\begin{equation} g(x) = (1/2\pi i) \int _{C} M(s) x^{- s}ds,
\nonumber \end{equation} where $C$ means a special contour on a
complex plane and compare an expression thus obtained with
the definition of the Fox $H$-function (Fox, C. 1961)\cite{fox61}
in a form of contour integral also \begin{eqnarray} \label{fox}
H_{pq}^{mn}\left[ x \left| \begin{array}{ccc} (a_{1}, A_{1}),&
\ldots , & (a_{p}, A_{p})\\ (b_{1},B_{1}),& \ldots, & (b_{q},
B_{q})\end{array} \right.  \right] = \nonumber \\[0.2 cm]
\frac{1}{2\pi i} \int _{C} \frac{\prod_{j=1}^{m}\Gamma(b_{j} +
B_{j}s)\prod_{j=1}^{n}(1 - a_{j} -
A_{j}s)}{\prod_{j=n+1}^{p}\Gamma(a_{j} +
 A_{j}s)\prod_{j=m+1}^{q}(1 - b_{j} - B_{j}s)}  x^{- s}ds.
 \nonumber \end{eqnarray}
As a result we get an expression for stable density in terms of
 the Fox $H$-function (Schneider, W. 1986)\cite{sch86}
 \begin{equation} g(x,\alpha ,\rho)=H_{22}^{11}\left[ x \left|
\begin{array}{cc} (-1/\alpha, 1/\alpha),& (-\rho, \rho)\\
(-1,1),&(-\rho, \rho)\end{array} \right. \right] \end{equation}

Using elementary properties of the Fox $H$-function we can show
that the stable densities for any values of parameters
$\alpha$ and $\rho$ are solutions of some integro-differential
equations with the integration and differentiation operators
in general of the fractional order.

If the Fox $H$-function (\ref{fox}) has rational parameters
$A_1,\ldots, A_p,B_1,\ldots,B_q$ then it degenerates to the Meyer
$G$-function that satisfies the ordinary differential equa\-tion
with polynomial coefficients of the order $\max (p,q)$. This
differential equation has 2 (if $p \not= q$) or 3 (if $p = q$)
singular points.  Thus if the parameters $\alpha, \rho$ are
rational numbers $\alpha = M/N, \ \rho = L/M, \quad (M,N)= 1,$ then
the stable densities are solutions of the usual differential
equations of the order $\max (M - 1, N - 1)$ and are expressed in
terms of the Meyer $G$-function (Zolotarev, V.M. 1994)\cite{zol94}:
\begin{eqnarray} g(x, M/N,L/M) = (2 \pi)^{L - (1/2)(M + N)}
(MN)^{1/2} x^{-1} \times \nonumber \\ G_{N+L-2,
M+L-2}^{M-1,L-1}\left(\frac{x^{M}N^{N}}{M^{M}} \left|
\begin{array} {cc} \frac{1}{N}, \ldots , \frac{N-1}{N},&
\frac{1}{L}, \ldots, \frac{L-1}{L}\\[0.1 cm] \frac{1}{M}, \ldots,
\frac{M-1}{M},& \frac{1}{L}, \ldots,
\frac{L-1}{L}\end{array}\right. \right).  \end{eqnarray}

As we see the properties of stable probability measures depend
essentially on whether critical indexes are rational or
irrational. In the modern theory of phase transitions the models
with rational indexes are well known e.g. the Ising model, minimal
models in 2-dimensional conformal statistical mechanics etc.

{\em E. Many-dimensional stable random variables and
phase transitions.}

The many-dimensional stable random
variables are very important from point of view of
applications. The L\'evy spectral measure for the
many-dimensional random variable has much more complicated form
than that for the one-dimensional case when as we know it is just a
power function. And this is one of the main difficulties in the
many-dimensional case.

First of all let us give a short review of stable random variables
in ${\mathbb R}^k$ \cite{jam93} \cite{hmv94}.

A random variable $X$ is called stable if there exist a sequence
of independent and equally distributed random variables
$\lbrace X_{n}\rbrace$ which after normalization it by the linear
operators $A_{n}\in GL({\mathbb R}^k)$ and centering by vectors
$b_{n} \in {\mathbb R}^k$ converge by distribution to a random
variable $X$, \begin{equation} A_{n}
\sum_{j=1}^{n} X_{j} + b_{n} \Rightarrow X.  \end{equation}

A full probability measure $\mu$ corresponding to the random
variable $X$  is stable if and only if there exist an operator
$B$ and a vector-valued function $b(t)$ of such a type that for
all $t > 0$ \begin{equation} \label{opl}
\mu^{t} = t^{B}\mu \ast \delta (b(t)), \end{equation} or which is
the same \begin{equation} \label{dud} \hat \mu^{t} (y)= \hat \mu
(t^{B^{*}}y) \exp (i\langle y,b(t)\rangle) \end{equation}

(Sakovich, G.N. 1961,1965\cite{sak61}\cite{sak65};
Sharpe, M. 1969\cite{sha69})

In accordance with this result an infinitely divisible measure
$\mu = [a,R,M]$ is stable if and only if for all $t > 0$
\begin{eqnarray}
[ta, tR, t \cdot M] = [a(t), t^{B}Rt^{B^{*}}, t^{B}M],
\\ \mbox{where} \qquad a(t) = b(t) + t^{B}a + \int
\left[\frac{x}{1 + \Vert x \Vert ^{2}} - \frac{x}{1 + \Vert t^{-
B}x \Vert ^{2}}\right] t^{B} M(dx).\nonumber \end{eqnarray}

The vector  $b(t)$ that appears in this formulae  has the
following form :
\begin{equation} \label{bnm} b(t) = t \int_{1/t}^{1} v^{-B}dv
\quad d, \end{equation} where $d$ is some constant
vector in ${\mathbb R}^k$.  (It is possible that this formula is
new, we have not succeed to find it in the available literature).

A set \begin{equation} E(\mu) =
\lbrace B \in GL({\mathbb R}^k) \vert \ \mu^{t} = t^{B} \mu
\ast \delta (b(t)) \rbrace \end{equation} we shall call a set of
exponents for the stable measure $\mu$ under consideration.

A set
\begin{equation}
S(\mu) = \lbrace A \in GL({\mathbb R}^k) \vert \ A \mu = \mu
\ast \delta (a) \ \mbox{for some} \quad a \in {\mathbb
R}^k\rbrace \end{equation} we shall call a symmetry group for
the measure $\mu$.

A full stable measure $\mu$ has one and only one exponent if
its symmetry group is finite.

Let $\Sigma (B)$ be a spectrum of the operator $B$. We can show
that \begin{equation} \Sigma (B) \subset \lbrace z
\in {\mathbb C}^k \vert \ \Re z \geq 1/2 \rbrace .\end{equation}

Let $\mu$ be a full stable measure with an exponent $B$.
Let us present a minimal polynom of the operator $B$ in a form of
product of polynomials $g_{1},g_{2}$ of such a type that
a real part of zeros for the polynom $g_{1}$ is equal to  $1/2$
and a real part of zeros for the polynom $g_{2}$ is strictly
greater than $1/2$.  Then we can present the measure $\mu$
in a form \begin{equation} \mu = \mu_{1} \ast \mu_{2},
\end{equation} where $\mu_{1}$ is the measure of normal random
variable and the measure $\mu_{2}$ has not any normal component.
Moreover \begin{eqnarray*} ker \ g_{1}(B) = supp \ \mu_{1}, \ ker
\ g_{2}(B) = lin(supp \ \mu_{2}), \ \mbox{and}\\ {\mathbb R}^k
= supp \ \mu_{1} \oplus lin(supp \ \mu_{2}).  \end{eqnarray*}

If a full stable measure $\mu$ is not normal then its
moments \begin{equation}
\alpha_{p} = \int \Vert x \Vert ^{p} \mu(dx) \nonumber
\end{equation} satisfy the following inequalities :
\begin{eqnarray} \alpha_{p} < \infty \quad
\mbox{for} \ 0 < p < 1/\Lambda, \\ \alpha_{p} = \infty \quad
\mbox{for} \  p > 1/\Lambda, \ \mbox{where}\\ \Lambda = max
\lbrace \Re x \vert \ x \in \Sigma (B) \rbrace.  \end{eqnarray}

If $1 \not\in \Sigma (B)$ then there exist a vector $a \in
{\mathbb R}^k$ which depends on $B$ and is of such a type that
the measure $\nu = \mu \ast \delta(a)$ is strictly stable i.e.
\begin{displaymath}
\nu^{t} = t^{B}\nu.
\end{displaymath}

Each stable measure in ${\mathbb R}^k$ is absolutely
continuous with respect to the Lebesgue measure on ${\mathbb
R}^k$ and its density has derivatives of all orders.

\section {Example: 2--parametric scaling theory of phase
transitions.}

Now let us analyze a simple example of application for
the theory put forward above in assumption that the random variable
characterizing macroscopic body takes its values in ${\mathbb
R}^2$.

Let the stable random variable describing macroscopic body under
phase tran\-si\-ti\-on has an exponent of the following form
\begin{equation} B =
\left( \begin{array}{cc} 1/\alpha_{1} & 0 \\ 0 & 1/\alpha_{2}
\end{array}\right) , \ 0 < \alpha_{1} \leq 2, \ 0 < \alpha_{2}
\leq 2.  \end{equation}

We have to study cases $1 \not\in \Sigma (B)$ and $1 \in
\Sigma (B)$ separately.

(1) At first let be  $\alpha_{1} \not= 1,
\alpha_{2} \not= 1$.

Under this condition we have according to the formula (\ref {bnm})
\begin{equation} b(t) =
\left(\frac{t^{1/\alpha_{1}} - 1 }{ 1/ \alpha_{1} - 1} d_{1},
\frac{t^{1/\alpha_{2}} - 1 }{ 1/ \alpha_{2} - 1} d_{2}\right),
\end{equation} where $d_{1}, d_{2}$ are some constants.

Let us take a measure $\mu$ in the form
\begin{equation} \label{puk}
\mu = \nu \ast \delta\left( - \frac{d_{1}}{1/\alpha_{1} -
1}\right) \ast \delta \left( - \frac{d_{2}}{1/\alpha_{2} -
1}\right).  \end{equation} Then substituting this expression for
the measure in (\ref{opl}) we can con\-vince our\-sel\-ves that the
appropriate random variable is strictly stable i.e. $\nu^{t} =
t^{B} \nu$.  It is worthwhile to point out here that a convolution
of a probability measure with a degenerate one (i.e. the
delta--function) is equivalent to changing a frame to reference
of the random variable. Thus in accordance with the formula
(\ref{puk}) when we change a frame of reference of the random
variable then with assumption $1 \not\in \Sigma(B)$ we can always
make the random variable strictly stable that we shall suppose
further.

According to (\ref{dud}) the logarithm of characteristic function
$\tilde \mu (y_{1}, y_{2})$ of the strictly stable random variable
satisfy the equation
\begin{equation} \label{rab} q \tilde \mu
(y_{1}, y_{2}) = \tilde \mu (q^{1/\alpha_{1}} y_{1}, \
q^{1/\alpha_{2}} y_{2}), \ q > 0.  \end{equation}

This functional equation has general solution (see for example
\cite{sak61}) \begin{equation} \label{bar} \tilde \mu (y_{1},
y_{2}) = (y_{1}^{\alpha_{1}} + y_{2}^{\alpha_{2}}) \ \tilde \nu (
y_{1}^{\alpha_{1}} / y_{2}^{\alpha_{2}}), \end{equation} where
$\tilde \nu(y)$ is arbitrary function which in the case under
consideration is analytical with respect to an appropriate
local variable. Let us continue the analytical function
$\tilde \mu (y_{1}, y_{2})$ in a complex plane and introduce
generally accepted in physics notations:  $y_{1} = t$ is
a dimensionless temperature, $y_{2} = h$ is a dimensionless field,
$\tilde \mu (y_{1}, y_{2}) = \Phi(t,h)$ is a singular part of
thermodynamic function under phase transition. In new notations
the equation (\ref{rab}) is nothing else than the equation of
scale invariance for singular part of thermodynamic function
under phase transition which is well known in physics.
The thermodynamic function itself according to (\ref{bar})  takes
such a form:
\begin{equation} \label{rad} \Phi (t,h) = (t^{\alpha_{1}} +
h^{\alpha_{2}}) \Psi (t^{\alpha_{1 }} / h^{\alpha_{2}} ).
\end{equation}

Now let us define the critical indexes of the heat
capacity $C$, the order parameter $\eta$ and
the susceptibility $\chi$ which we can express through the
thermodynamic func\-ti\-on in such a way:  \begin{equation}
\label{arn} C = - \frac{\partial ^{2} \Phi}{\partial t^{2}}, \quad
\eta = - \frac{\partial \Phi}{\partial h}, \quad \chi = -
\frac{\partial ^{2} \Phi}{\partial h^{2}}.  \end{equation}

Let us consider separately the cases of weak and strong
fields.

(*) The weak fields, $t^{\alpha_{1}} > h^{\alpha_{2}}$. In this
case  \begin{equation} \label{aks} \Phi (t,h) =
t^{\alpha_{1}} f(h/ t^{\alpha_{1} /  \alpha_{2}}, sign \ t),
\end{equation} where $f(x, \pm)$ are
real-analytical functions with the
following power expansions:  \begin{eqnarray} f(x,+) =
1 + f_{1}^{+} x^{2} + f_{2}^{+} x^{4} + \ldots \\ f(x,-) = 1 +
f_{1}^{-} x + f_{2}^{-} x^{2} + \ldots .  \end{eqnarray}

We define the critical indexes $\alpha, \beta, \gamma$ in the
range of weak fields by such a way:
\begin{equation} \label{cri} C \sim \vert t \vert^{-\alpha}, \quad
\eta \sim \vert t \vert^{\beta} \ \mbox{for} \ t < 0, \quad \chi
\sim \vert h \vert^{- \gamma}\ \mbox{for}\ t > 0.  \end{equation}

(**) The strong fields, $t^{\alpha_{1}} <
h^{\alpha_{2}}$.  In this case \begin{equation}
\label{tok} \Phi (t,h) = h^{\alpha_{2}} g(t/ h^{\alpha_{2} /
\alpha_{1}}), \end{equation} where $g(x)$ is real-analytical
function with the following power expansion:
\begin{equation} g(x) = 1 + g_{1} x + g_{2} x^{2} +
\ldots .  \end{equation}

We define the critical indexes $\epsilon, \delta$ in
the range of strong fields by such a way:
\begin{equation} \label{tic} C \sim h^{-\epsilon}, \quad
\eta \sim h^{1/\delta}.  \end{equation}

Now substituting (\ref{aks}) and (\ref{tok}) in (\ref{arn}) and
comparing the results of sub\-sti\-tu\-ti\-on with (\ref{cri}) and
(\ref{tic}) appropriately we get the following expression for
the critical indexes defined above in terms of the
characteristic exponents:
\begin{eqnarray} \label{che} \alpha = 2 - \alpha_{1}, \ \beta =
(\alpha_{2} - 1)\alpha_{1} / \alpha_{2}, \ \gamma = (2 -
\alpha_{2})\alpha_{1}/ \alpha_{2},\nonumber \\ \epsilon = (2 -
\alpha_{1})\alpha_{2} / \alpha_{1}, \ \delta = 1 / (\alpha_{2} -
1).  \end{eqnarray}

Analyzing asymptotic expansions for the correlation function of
order parameter
\begin{equation} \label{cor} G(r) = \langle
\eta(r) \eta(0) \rangle \sim r^{-\sigma} \exp(- r / r_{c}(t,h)), \
r\to \infty, \end{equation} where $r_{c}(t,h)$ is the correlation
radius, physicists introduce three more critical indexes $\nu, \mu,
\zeta$ :
\begin{equation} r_{c}(t,0) \sim \vert t \vert^{-\nu}, \
r_{c}(0,h) \sim h^{-\mu},\ \zeta = \sigma - d + 2.  \end{equation}
We can easily get indexes  $\nu,\ \mu$ from the equation of scale
invariance of the singular part of thermodynamic function
(\ref{rab}) in a form \begin{equation}
\label{muk} u^{d}\Phi (t,h) = \Phi (tu^{1/\nu},\
hu^{1/\mu}),\end{equation} where $u$ is a unit of measurement of a
linear size of macroscopic body and $d$ is a dimensionality of a
space containing the body. Substituting in this
equation the thermodynamic function in the form (\ref{rad})
we find that \begin{equation} \label{nuk} \nu = \alpha_{1}/ d, \
\mu = \alpha_{2}/ d.\end{equation}

Under assumption $r_{c} = \infty$ in the asymptotic expression
(\ref{cor}) for the correlation function of order parameter
we can show easily that \begin{equation} \label{zet} \zeta =
d(\alpha_{2} - 2) /\alpha_{2} + d .\end{equation}

The formulae (\ref{che}), (\ref{nuk}), (\ref{zet}),
express the critical indexes of phase tran\-si\-ti\-ons in terms of
cha\-rac\-te\-ris\-tic exponents and reproduce completely
results of so called the 2-parametric scaling theory of phase
transitions.

We want to draw attention of the reader to some
preferred values of the cha\-rac\-te\-ris\-tic exponents.

For the characteristic exponent $\alpha_{1} \in ( 0, 2]$
the value $\alpha_{1} = 2$ is singled out. Under this value of
the characteristic exponent the domain of attraction or, which is
the same, the universality class of phase transition is the
largest possible one and therefore we should wait for the value
$\alpha_{1} = 2$ in majority of phase transitions.

According to definition  $\alpha_{2} \in ( 0,
2].$ However if we make in addition a very natural assumption that
the quantity $\sigma = d - 2 + \zeta = 2d(\alpha_{2} - 1) /
\alpha_{2}$ which appears in the expression for the correlation
function (\ref{cor}) is non-negative then we should accept that
$\alpha_{2} \in (1,2]$.  If $\alpha_{2} \to 1 + 0$ then  $\sigma
\to +0$ and therefore in a critical range where $r_{c} = \infty$
and $G(r) \sim r^{-\sigma}$ the correlations of the order
parameter will decrease very slowly. This circumstance singles out
the value $\alpha_{2} = 1$.  (We shall not discuss here an
interesting case when $\alpha_{2} \in [0,1]$. In this case
the correlations of order parameter in critical range should
increase with distance.)

Let  us apply these considerations to analysis of the
critical indexes of phase transitions which we have
in fact in different cases.

Let us suppose that $d \to \infty$ as it is indeed in the mean
field theory.  In this case limit values of the critical
indexes characterizing the correlation function do not depend
on the characteristic exponents:  \begin{equation} \nu = 0, \  \mu = 0, \ \zeta = -
\infty.  \end{equation}

If in addition we assume that
\begin{equation}
\alpha_{1} = 2, \ \alpha_{2} = 4/3,
\end{equation}
then  other critical indexes take the so called ``classical"
values
\begin{equation} \label{hem} \alpha = 0, \
\beta = 1/2, \ \gamma = 1,\ \epsilon = 0, \ \delta = 3.
\end{equation}

Let us consider  $d = 3$. If in addition we assume that
\begin{equation} \alpha_{1} = 2, \ \alpha_{2} =
6/5, \end{equation} then the critical indexes take rational
values
\begin{equation} \label{hes} \alpha = 0, \ \beta = 1/3, \ \gamma =
4/3,\ \epsilon = 0, \ \delta = 5, \ \nu = 2/3, \ \mu = 2/5, \
\zeta = 0, \end{equation} which describe well enough
numerous experimental data. However the following values of
the characteristic exponents lead to better agreement with an
experiment :  \begin{equation} \alpha_{1} = 1,89 \pm 0,05, \
\alpha_{2} = 1,210 \pm 0,005, \end{equation}

(2) Now let be  $\alpha_{1} = 1,\ \alpha_{2} \not= 1$.

Under this assumption we have according to the formula
(\ref{bnm}), \begin{equation} b(t) = \left( t \ln t \ d_{1},
\frac{t^{1/\alpha_{2}} - t} {(1/\alpha_{2}) - 1} \ d_{2} \right)
,\end{equation} where $d_{1}, \ d_{2}$ are some constants.
We may put $d_{2} = 0$ if we change in appropriate way a point of
reference.  Assuming that we can present the probability measure
$\mu $ in the following form
\begin{equation} \mu = \nu \ast \lambda,\end{equation} where
$\nu $ is a strictly stable measure and \begin{equation} \hat
\lambda (y_{1}, y_{2}) = -id_{1}y_{1} \ln \vert y_{1} \vert .
\end{equation}

Now similarly to the previous case we get for the logarithm
of cha\-rac\-te\-ris\-tic function the following expression:
\begin{equation} \hat \mu
(y_{1}, y_{2}) = -i d_{1} y_{1} \ln \vert y_{1} \vert + (y_{1} +
y_{2}^{\alpha_{2}}) \rho(y_{1}/ y_{2}^{\alpha_{2}}),
\end{equation} where $\rho(z)$ is an analytic function with
respect to some local variable. Con\-ti\-nu\-ing analytically
the function $\hat \mu (y_{1}, y_{2})$ in a complex plane and
ascribing a special physical meaning to its independent
variables we obtain from $\hat \mu (y_{1}, y_{2})$ a
singular part of the thermodynamic function in a vicinity
of phase transition.

If e.g. we assume that $\alpha_{2} =
16/15, \ y_{1} = t^{2}, \ \mbox{where} \ t$ is the dimensionless
temperature and $y_{2} = h$ is dimensionless field then the
function $\hat \mu (y_{1}, y_{2})$ is trans\-for\-med to a
singular part of the thermodynamic function of the
exactly solvable Ising model
\begin{equation} \Phi (t,h) = ct^{2} \ln \vert t \vert + (t^{2} +
h^{16/15}) \Psi (t^{2}/ h^{16/15}) , \end{equation} $c$ is some
constant. In order to see that it is enough to consider limit
cases
\begin{eqnarray} \Phi (t,h) \simeq t^{2} (c_{1} \ln \vert
t \vert + c_{2}) + c_{3} h t^{1/8}, \ t > h^{8/15} \\ \Phi (t,h)
\simeq c_{4} h^{16/15} + t^{2}(c_{5} \ln \vert t \vert + c_{6}), t
< h^{8/15} , \end{eqnarray} where $c_{i}, i = 1,\ldots,6$ are some
constants. It is well known classical result
(On\-sa\-ger, L. 1944, 1947 \cite{ons44}, \cite{ons47}).

In other exactly solvable models of statistical mechanics the
characteristic ex\-po\-nent $\alpha_{2}$ is a constant or a
rational function of the characteristic exponent $\alpha_{1}$.

We obtain similar results for the macroscopic bodies in state of
phase transition described by random variables with the following
exponents \begin{equation} B = \left(\begin{array}{cc} 1/\alpha &
0 \\ 0 & 1/\alpha \end{array} \right), B = \left(\begin{array}{cc}
1/\alpha & 0 \\ \beta & 1/\alpha \end{array} \right), B =
\left(\begin{array}{cc} 1/\alpha & \beta \\ \beta & 1/\alpha
\end{array}\right).  \end{equation}

\section {Conclusion.}

In the paper we have given a new formulation of statistical
mechanics and built on this basis a theory of phase transition.
According to this approach the random variables describing
macroscopic bodies in statistical mechanics are infinitely
di\-vi\-sible that is just mathematical expression of an
atomic-molecular structure of the matter. In a state of
phase transition these random variables are stable that
reflects an invariance of the thermodynamic functions of
macroscopic bodies with respect to transformations of the
renormalization semigroup.

We should point out some questions which have been studied
but have not presented here:  different examples of
computation of critical indexes, discussion of limit cases
(the mean field theory, the Landau theory, the Lee and Yang
theory etc.), the non-ergodic systems and spin
glasses, the non-equilibrium statistical mechanics,
an interconnections with microscopic theories. We are going to
publish them elsewhere.

\section* {Acknowledgments}

The author express his gratitude to Kochubei A.N. for his
intertest to this work and the reference \cite {sch86} and to
Enolskii V.Z. for his help to get the book \cite{jam93}.

\end{document}